\documentclass[twocolumn,pra]{revtex4}
\usepackage{amsmath,amsfonts}
\usepackage{graphicx}

\def\be{\begin{equation}}
\def\ee{\end{equation}}
\def\bea{\begin{eqnarray}}
\def\eea{\end{eqnarray}}
\def\nn{\nonumber}

\begin{document}
\title{\bf Interference-assisted detection of dark photon
using atomic transitions}

\author{V.V. Flambaum$^{1,2}$, I.B. Samsonov$^{1}$, H.B. Tran Tan$^{1}$}
\affiliation{$^1$School of Physics, University of New South Wales,
Sydney 2052,  Australia,}
\affiliation{$^2$Johannes
Gutenberg-Universit\"at Mainz, 55099 Mainz, Germany}

\begin{abstract}
Dark photon is a massive vector particle which couples to the
physical photon through the kinetic mixing term. Such particles,
if exist, are produced in photon beams and, in particular, in
laser radiation. Due to the oscillations between the physical
photon and the dark photon, the latter may be, in principle,
detected in the light-shining-through-a-wall experiment. We
propose a variant of this experiment where the detection of dark
photons is based on the atomic transitions. The key feature of
this scheme is that the detection probability is first order in
the coupling constant due to the interference term in the photon
and dark photon absorption amplitudes. We expect that such
experiment may give new constraints on dark photon coupling
constant in the mass region $10^{-3}< m < 10^{-2}$ eV.
\end{abstract}

\maketitle

\section{Introduction}

The origin and nature of dark matter is one of the great enigmas in contemporary physics. Although there exist many cosmological observations indicating the presence of the dark matter, its properties remain essentially unknown due to the feebleness of its interactions with baryonic matter. The most well-motivated candidates for dark matter include the weakly interacting massive particle (WIMPs), the axion and the dark photon. In this paper, we will focus on the last option.

The dark photon is a massive vector particle which originates from a conjectured new $U(1)$ symmetry extension of the Standard Model. This new massive vector particle of mass $m$ couples to the massless photon through the kinetic mixing term with a coupling constant $\chi$. There are strong constraints on the parameters $m$ and $\chi$ originating from laboratory experiments and astrophysical arguments, see, e.g., Refs.\ \cite{Raggi:2015yfk,Irastorza:2018dyq,Jaeckel:2013ija}.

Assuming that the dark matter consists of the dark photon, it is natural to look for manifestations of the latter in cosmic rays coming either from the Sun \cite{Arik:2008mq} or from the galactic halo using different types of detectors, see, e.g., \cite{Nelson:2011sf} for a review. Alternatively, one can assume that the dark photons (analogously to the axions) are produced in laboratory in laser beams due to the photon-dark photon oscillations. The corresponding experiments are usually referred to as the light-shining-through-a-wall (LSW) ones, among whom are ALPS \cite{Ehret:2007cm,ALPSII1,ALPSII2}, BMV \cite{Robilliard:2007bq}, GammeV \cite{Chou:2007zzc}, LIPSS \cite{Afanasev:2008fv}, OSQAR \cite{Pugnat:2013dha,Pugnat:2007nu,Ballou:2015cka} and BFRT \cite{Cameron:1993mr,Ruoso}. 

The detection of cosmic dark photons and axions is (theoretically) facilitated by the observation that they should be produced in abundance inside stars \cite{An:2013yfc,An:2013yua,An:2014twa}, but the corresponding detectors should be of broad-band type as the energies of these particles are unknown. This reduces the sensitivity of such detectors as compared with the resonance-based detectors; the latter may be employed in the LSW-type experiments since they are aimed at observe the laboratory produced dark photons.

In our recent paper \cite{Tan:2018tcs}, we proposed an improvement of the classical LSW experiment with enhanced sensitivity to axion-like particles. The main idea was to use a semi-transparent wall which allows a fraction of the photon beam to reach the detector. In this case, the axion signal may be identified with the interference term between the photon and axion resonant absorption amplitudes in the detection medium. The necessary condition for this scheme to work is the coherence between the axion and photon beams which may be easily satisfied for existing experiments assuming that the mass of the axion is much smaller than its energy. The advantage of this detection scheme becomes evident if one notices that this interference term is first order in the axion coupling constant while the standard axion absorption probability is of the second order in this coupling. 

In this paper, we develop a similar idea for dark photon LSW experiments. Indeed, if we have a source of coherent beams of the photons and dark photons, the latter may be detected by observing the interference of the absorption of these particles. As a result, the signal due to dark photons becomes first order in the coupling constant. Although this idea is very similar to the axion interference-assisted detection scheme \cite{Tan:2018tcs}, there are two important features.

First, the mechanism for producing dark photons is different from that for axions. Recall that the axions may be born in the laser beam passing through a (strong) magnetic field due to the $a {\bf E}\cdot {\bf B}$ interaction. This magnetic field provides a reversal which allows the enhancement of the axion signal and the elimination of systematics. In the dark photon case, the oscillation between the photon and dark photon happens independently of the external fields because of the kinetic mixing term $F^{\mu\nu}F'_{\mu\nu}$. Thus, there is no effective control over the dark photon production. However, the dark photon production probability may be reduced when the photon beam passes through a medium \cite{An:2013yfc,An:2013yua,An:2014twa}, while the resonant cavity may enhance the dark photon beam \cite{Ahlers:2007rd,Ahlers:2007qf}.

Second, the selection rules for the absorption of axion and dark photon particles in atoms are different as they are described by the pseudoscalar and vector fields, respectively. Therefore, it is natural to expect that the detectors for these particles should use different atomic (or molecular) transitions. In Ref.\ \cite{Tan:2018tcs}  we demonstrated that it is advantageous to employ M1 transitions in molecules or in crystals for detecting axions. For dark photon, however, E1 transitions are more beneficial since the corresponding selection rules are similar to the usual photon absorption and E1 transitions generally have the largest amplitude. 

The rest of the paper is organized as follows. In Sect.\ \ref{Sec2}, we describe the scheme of the experiment for detecting dark photons using atomic transitions based on the interference between the photon and dark photon beams. In Sect.\ \ref{Theory}, we develop a theoretical background for the proposed experiment. In this section, we start with a review of different representations of the dark photon Lagrangian which are suitable either for description of the photon-dark photon oscillations or for detection of dark photons using atomic transitions, then we compute the oscillation probability between the dark proton and photon in medium, as well as derive the amplitudes of resonant absorption of dark photons with E1 atomic transitions. In Sect.\ \ref{Num}, we present numerical estimates for the proposed experiment and compare them with the results of other experiments. Sect.\ \ref{Disc} is devoted to a discussion of the results. In Appendix, we give the details of calculation of the index of refraction of the detecting medium which are needed for correct estimates of the expected dark photon signal.

Throughout this paper we use natural units in which $c=\hbar=1$.

\section{Proposed experimental setup}
\label{Sec2}

\begin{figure}[htb]
\begin{center}
\includegraphics[width=8cm]{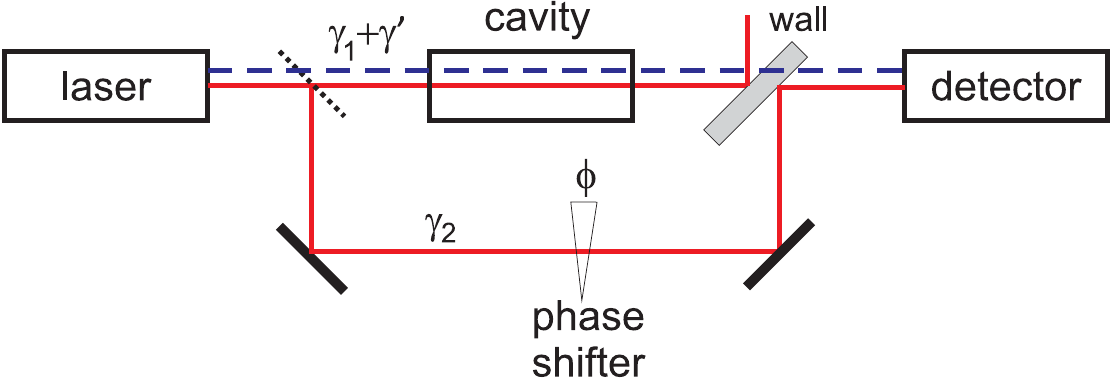}
\end{center}
\caption{Experiment setup.
Dashed (blue) line corresponds to the dark photon beam $\gamma'$ while the solid (red) lines represent the photon beams $\gamma_1$ and $\gamma_2$.}
\label{fig1}
\end{figure}

The schematic design of our proposed experiment is presented in
Fig.\ \ref{fig1}. In this scheme, a laser produces a coherent mixture
of photons and dark photons. The photon beam is split by a mirror
into the main beam $\gamma_1$ (in the same direction as the original beam) and the reference beam $\gamma_2$
(lower path). The main beam is passed through a resonant cavity
which amplifies the dark photon signal (see, e.g., Appendix B in Ref.\
\cite{Ahlers:2007rd}). The photons emerging from the cavity are then absorbed or deflected away by
 a wall which makes negligible suppression of the dark photon
beam. The dark photon beam traversing the wall meets the reference beam in the
detector, where they are coherently absorbed. The coherence of the dark and normal photons means that there is an interference between the dark and normal photon absorption amplitudes. The interference term depends on the phase difference $\phi$ between the reference photon beam and the dark photon beam. Thus, the dark photons may be detected by measuring the variation of the signal in the detector when the length of the path of the reference beam is changed.

The detection medium may be a gas or other media which
consist of atoms with transitions in resonance with the laser
beam. We denote the ground state of the atoms in the detector by
$A$ and the excited state by $B$, the transition energy is
$\omega_{BA}\equiv \omega_B - \omega_A=\omega$, where $\omega$ is the energy of the
photons and dark photons. The signal from the dark photon may be
observed by measuring the variation in the population of atoms in
the excited state $B$ when the phase of the reference beam is
varied.

\section{Theory}
\label{Theory}
We start this section with a short review of different but equivalent Lagrangian formulations of the dark photon model which will be further applied in deriving the dark photon production and absorption rate in medium. This will allow us to estimate the signal due to dark photon detection and the corresponding signal-to-noise ratio (SNR).

\subsection{Dark photon Lagrangian}
The dark photon ($\gamma'$) is described by a massive vector field $A'_\mu$ which
weakly couples to the physical photon ($\gamma$) $A_\mu$ through the kinetic
mixing term with the coupling constant $\chi$. The dark photon extension of the QED
Lagrangian reads
\bea
 {\cal L} &=& -\frac14 F_{\mu\nu}F^{\mu\nu} - \frac14 F'_{\mu\nu}F'^{\mu\nu}
 -\frac\chi2 F_{\mu\nu} F'^{\mu\nu}  \nn\\&&
 + \frac{m^2}{2} A'_\mu A'^\mu +e J^\mu_{\rm em} A_\mu\,,
\label{1}
\eea
where $F_{\mu\nu} = \partial_\mu A_\nu - \partial_\nu
A_\mu$ and $F'_{\mu\nu} = \partial_\mu A'_\nu - \partial_\nu
A'_\mu$ are the field strengths for the physical and dark photons,
respectively. Here $m$ is the mass of the dark photon and $J^\mu_{\rm
em}$ is the matter current.

In the Lagrangian (\ref{1}), it is convenient to eliminate the
mixing
term $\frac\chi2 F_{\mu\nu} F'^{\mu\nu}$ by performing a change of
fields. There are two basic ways to do that (see e.g.\ \cite{Jaeckel:2013ija} for a review):

\begin{itemize}
\item[i)] The kinetic mixing may be turned into the mass mixing using
the change of fields
\be
A'_\mu \to A'_\mu - \chi A_\mu\,.
\ee
Upon this change of fields the Lagrangian (\ref{1}) becomes (modulo
$O(\chi^2)$ terms)
\bea
{\cal L}&=& -\frac14 F_{\mu\nu}F^{\mu\nu} - \frac14 F'_{\mu\nu}F'^{\mu\nu}
 \nn\\&&
 + \frac{m^2}{2} (A'_\mu A'^\mu -2 \chi A'^\mu A_\mu) +e J^\mu_{\rm em} A_\mu\,.
\label{3}
\eea
In this representation, only the field $A_\mu$ interacts with
matter, but the fields $A_\mu$ and $A'_\mu$ do not describe the
stationary mass states. As a result, in vacuum, there are oscillations between
these fields with probability \cite{Ahlers:2007rd,Ahlers:2007qf}
\be
P_{\gamma\to\gamma'}(l) =
4\chi^2
 \sin^2\frac{l m^2}{4\omega}\,,
\label{4}
\ee
where $\omega = \sqrt{{\bf k}^2+m^2}$ is the energy of the dark photon
with momentum ${\bf k}$ and $l$ is the distance the photon
travels. In Fig.\ \ref{fig1}, $l$ may be identified with the distance between the laser and the detector.

\item[ii)] The mixing term in Eq.\ (\ref1) may be fully removed in
the first order in the coupling constant $\chi$ by the shift of
the photon field
\be
A_\mu \to A_\mu - \chi A'_\mu\,.
\ee
However, upon this change of fields, the Lagrangian (\ref{1}) acquires the
interaction term of dark photons with the matter fields:
\bea
{\cal L}&=& -\frac14 F_{\mu\nu}F^{\mu\nu} - \frac14 F'_{\mu\nu}F'^{\mu\nu}
 + \frac{m^2}{2} A'_\mu A'^\mu \nn\\&&
 +e J^\mu_{\rm em} A_\mu-\chi e J^\mu_{\rm em} A'_\mu\,.
\label{6}
\eea
In this representation, there are no oscillations between the fields
$A_\mu$ and $A'_\mu$ since they describe stationary states in
vacuum.

\end{itemize}

We stress that the Lagrangians (\ref{3}) and (\ref{6}) describe
the same physics. However, the Lagrangian (\ref{6}) is more suitable
for computing atomic transitions due to the dark photon absorption
and emission.

\subsection{Dark photon oscillation in medium}
\label{Sec-osc}

Let us consider a non-magnetic medium characterized by a (complex) refractive
index $n$. A modification of the Lagrangian (\ref{3}) which describes the 
propagation of the coupled $\gamma-\gamma'$ system in medium reads
\bea
{\cal L}&=& -\frac14 F_{\mu\nu}H^{\mu\nu} - \frac14 A'_{\mu\nu}A'^{\mu\nu}
\nn\\&& + \frac{m^2}{2} (A'_\mu A'^\mu -2 \chi A'^\mu A_\mu)\,,
 \label{7}
\eea
where the tensor $H^{\mu\nu}$ has the following components
\be
H^{\mu\nu} =
\left(
\begin{array}{cc}
 0 & -  D_i \\
  D_j & -\varepsilon_{ijk}B_k
\end{array}
\right)\,.
\ee
Here $D_i$ are the components of the displacement field, ${\bf D} = (n^2-1){\bf
E}$, and $B_i$ are components of the magnetic field $\bf B$. The
Lagrangian (\ref{7}) implies the following equation of motion for
the vectors ${\bf A}$ and ${\bf A}'$:
\be
\left(
\begin{array}{cc}
-n^2 \frac{d^2}{dt^2 } + \nabla^2 & -\chi m^2 \\
-\chi m^2 & -\frac{d^2}{dt^2} + \nabla^2 - m^2
\end{array}
\right)
\left(
\begin{array}{c}
{\bf A} \\ {\bf A}'
\end{array}
\right) =0\,.
\label{8}
\ee

We look for a solution of Eq.\ (\ref{8}) in the form of plane
waves along the $z$ direction with frequency $\omega$,
$({\bf A}, {\bf A}')= ({\bf A}_0, {\bf A}'_0) e^{-i(\omega t - k
z)}$. The two linearly independent solutions read
\be
\left(
\begin{array}{c}
1 \\ -X
\end{array}
\right) e^{-i(\omega t -k_1 z)}\,,\qquad
\left(
\begin{array}{c}
X \\ 1
\end{array}
\right) e^{-i(\omega t -k_2 z)}\,,
\label{2sol}
\ee
where
\be
X = \frac{\chi m^2}{m^2+ \omega^2 \Delta\varepsilon_{\rm r}}
\label{X}
\ee
is the medium-corrected coupling constant and $k_1,k_2$ are the
particle momenta
\be
k_1 = n \omega+O(\chi^2)\,,\qquad
k_2 = \sqrt{\omega^2 - m^2}+O(\chi^2)\,.
\ee
In Eq.\ (\ref{X}) we use the notation $\Delta \varepsilon_{\rm r} \equiv \varepsilon_{\rm r}-1 = n^2-1$, $\varepsilon_{\rm r}$ is the relative permittivity.
The photon and dark photon fields are linear combinations of the 
two solutions (\ref{2sol})
\begin{eqnarray}
{\bf A}(z,t) &=&{\bf A}_0
\left[
\frac{e^{-i(\omega t - k_1 z)}}{1+X^2}
+\frac{X^2e^{-i(\omega t - k_2 z)}}{1+X^2}
\right]\,,\\
{\bf A}'(z,t) &=&{\bf A}'_0
\left[
\frac{Xe^{-i(\omega t - k_2 z)}}{1+X^2}
-\frac{Xe^{-i(\omega t - k_1 z)}}{1+X^2}
\right]\,.
\end{eqnarray}
This solution corresponds to the boundary condition
$({\bf A}(0,t),{\bf A}'(0,t)) = ({\bf A}_0,0)$.

The probability of oscillation of a photon into dark photon
reads
\be
P_{\gamma\to\gamma'}(z) =
\frac{| {\bf A}' |^2}{|{\bf A}'_0|^2} = 4|X|^2 \sin^2\frac{ \Delta k z}{2}\,,
\label{10}
\ee
where
\be
\Delta k = k_1 - k_2 = n\omega -\omega
\sqrt{1-\frac{m^2}{\omega^2}}\,.
\ee

We assume that the energy of the photon and the dark photon are in
the eV region while the mass of the dark photon is in the sub-eV
region. This assumption is needed to have non-vanishing probability
of the dark photon production in atomic transitions.
In this case $m\ll\omega$, and Eq.\ (\ref{10}) simplifies
\bea
P_{\gamma\to\gamma'}(z) &\approx&
\frac{4\chi^2 m^4}{(m^2+\omega^2 {\rm Re}\Delta\varepsilon_{\rm r})^2
+(\omega^2 {\rm Im}\Delta\varepsilon_{\rm r})^2}
\nn\\&&\times
 \sin^2\left[\left(\frac{m^2}{4\omega}+(n-1)\frac\omega2\right)z\right]\,.
\label{11}
\eea
Here we substituted explicitly the medium-corrected coupling
constant (\ref{X}). We stress that the effect of the medium is significant when
$m^2\ll |\Delta\varepsilon_{\rm r}|\omega^2$. 

Eq.\ (\ref{11}) shows that any medium suppresses the production of dark photons. In order to maximize the photon amplitude one is to use a good vacuum for the photon beams. However, the medium correction of the coupling (\ref{X}) is important in the consideration of the resonant absorption of dark photons in detectors. The equivalent result was found in Refs.\ \cite{An:2013yfc,An:2013yua,An:2014twa}.

In conclusion of this subsection we also point out that one can
enhance the dark photon production amplitude by placing a
resonant cavity in the photon beam. As was demonstrated in Ref.\ 
\cite{Ahlers:2007rd}, a resonant cavity which allows for $N_{\rm
pass}$ reflections of the photon beam gives an enhancement of the
dark photon production amplitude by the factor $(N_{\rm
pass}+1)/2$. Thus, a laser beam with $N_{\gamma_1}$ photons per
second passing through the resonant cavity may produce
\bea
N_{\gamma'} &=& N_{\gamma_1} \frac{N_{\rm pass}+1}2
P_{\gamma\to\gamma'}\nn\\
&=&2 \chi^2 N_{\gamma_1} (N_{\rm pass}+1)
\sin^2 \frac{l m^2}{4\omega}
\label{Ndp}
\eea
dark photons per second. Here we used the expression (\ref{4}) for the dark photon production probability in vacuum. We will assume the estimate (\ref{Ndp}) in our further
calculations.

\subsection{Dark photon resonant absorption amplitude}

In vacuum, the propagation of the massive vector field is
described by a plain wave with wave-vector $\bf k $ and energy
$\omega$
\be
A'_\mu =\sum_{i=1}^3 A_0^i \epsilon^i_\mu
\sin(\omega t - {\bf k}{\bf r})\,.
\label{20}
\ee
Here $\epsilon^i_\mu$ are the polarizations vectors obeying the
conditions $\epsilon^i_\mu k^\mu=0$, $\epsilon^i_\mu
\epsilon^{i\mu}=-1$. In particular, when the dark photon
propagates along the $z$ axis we have $k^\mu=(\omega, 0, 0, \sqrt{\omega^2 -
m^2})$ and the polarization vectors may be chosen in the form
$\epsilon^{1\mu} = (0,1,0,0)$, $\epsilon^{2\mu} = (0,0,1,0)$,
$\epsilon^{3\mu} = m^{-1}(\sqrt{\omega^2 -
m^2},0,0,\omega)$. The polarization vectors $\epsilon^{1,2}_\mu$
correspond to the transverse polarization while $\epsilon^3_\mu$
is attributed to the longitudinal polarization.

The amplitude of the plain wave $A_0^i$ may be
expressed via the number density of dark photons with a given polarization
${\cal N}^i_{\gamma'}$, $A_0^i=\sqrt{8\pi \omega^{-1}{\cal N}^i_{\gamma'}}$.

Consider now an atom (or a molecule) in the initial state $A$ with energy $\omega_A$. Due
to the interaction with the dark photon field, the atom can be
resonantly excited to the state $B$ with energy $\omega_{B}$ such
that $\omega_{BA}\equiv \omega_{B}-\omega_A = \omega$. According
to the model (\ref{6}), the interaction Hamiltonian in the
relativistic form reads $H_{\rm int} = \chi e \gamma^\mu A'_\mu$,
where $\gamma^\mu$ are the Dirac matrices. E1 transition amplitudes
corresponding to the resonant absorption of the longitudinal ($L$) or
transverse ($T$) components of the field (\ref{20}) have the standard
forms
\begin{eqnarray}
M^{(T)}_{\gamma'} &=& -i e^{-i\omega_B t}t \chi e
\sum_{i=1,2}\sqrt{2\pi n_{\gamma'}^i\omega}
\langle B | \boldsymbol{\epsilon}^i {\bf r} | A \rangle\,,
\label{MT}\\
M^{(L)}_{\gamma'} &=&-i e^{-i\omega_B t}t \chi e m
\sqrt{2\pi n_{\gamma'}^3\omega^{-1}}
\langle B | z | A \rangle\,.
\label{ML}
\end{eqnarray}

Note that for $\chi=-1$ Eq.\ (\ref{MT}) describes the
standard E1 photon absorption amplitude,
\be
M_\gamma = M^{(T)}_{\gamma'}|_{\chi=-1}\,,
\label{Mgamma}
\ee
while Eq.\ (\ref{ML}) vanishes in the massless-dark-photon case.

Note that we have assumed that the mass of the dark photon is much
less than the atomic transition energy, $m\ll\omega$. In this
case, the absorption amplitude of the longitudinal component of
the dark photon is suppressed as compared with the transverse
component,
\be
|M^{(L)}_{\gamma'}|\ll |M^{(T)}_{\gamma'}|\,.
\ee
Therefore, in what follows we will focus on the transverse
components of the dark photon.

\subsection{Dark photon signal}
\label{Sec-signal}

The atoms in the target coherently absorb the dark photons in the main beam and the photons in the reference beam. The
total absorption amplitude is the sum of the
photon absorption amplitude (\ref{Mgamma}) and the dark photon absorption amplitude (\ref{MT}) with the vacuum coupling $\chi$ replaced by its in-medium counterpart (see Eq.\ \eqref{X})
\begin{subequations}
\label{Mtot}
\bea
M_{\rm tot}&=& M_{\gamma} + M^{(T)}_{\gamma'}\,,\\
M_{\gamma}&=&i e^{-i\omega t +i\phi} te \sqrt{2\pi \omega {\cal N}_{\gamma_2}}
d\,,\\
M^{(T)}_{\gamma'}&=&-i e^{-i\omega t }t\frac{\chi e m^2}{m^2 +\omega^2\Delta\varepsilon_{\rm r}} \sqrt{2\pi \omega {\cal N}_{\gamma'}} d\,,
\eea
\end{subequations}
where $d= \langle B| \boldsymbol{\epsilon}\cdot{\bf r} | A\rangle$ is
the electric dipole matrix element, $\boldsymbol{\epsilon}$ is the polarization of the dark and normal photons (for simplicity, we assume that these two polarizations are the same) and $\phi$ is the phase difference between the photon and the dark photon. This phase difference appears due to the difference in the path lengths of the photon and the dark photon, see Fig.\ \ref{fig1}. The time parameter $t$ in Eq.\ (\ref{Mtot}) is limited by the lifetime of the atom in the excited state $B$.

Squaring $M_{\rm tot}$, one obtains the absorption probability $\mathcal{P}$
\begin{equation}\label{prob}
    \mathcal{P}
    =|M_{\rm tot}|^2
    = \left|M_{\gamma}\right|^2+2{\rm Re}\left(\overline{M}_{\gamma}M^{(T)}_{\gamma'}\right)
    +O(\chi^2).
\end{equation}
The relative signal is defined as the ratio of the second term in the right-hand side in Eq.\ 
(\ref{prob}) to the first one,
\bea
&&S(\phi) = \frac{2{\rm Re}(\overline{M_\gamma}M_{\gamma'})}{|M_\gamma|^2} \nonumber\\
&&=4\chi^2\sqrt{\frac{N_{\gamma_1}(N_{\rm pass}+1)}{2N_{\gamma_2}}}
\left|
\sin\frac{lm^2}{4\omega}
\right|
{\rm Re}\frac{-e^{-i\phi}m^2}{m^2 + \omega^2 \Delta\varepsilon_{\rm r}}\,.~~~~~
\eea

In the experiment, the variation in the signal $S$ should be observed when the phase $\phi$ is varied from some vale $\phi_1$ to $\phi_2$. So, the observed quantity is $\Delta S = S(\phi_1)-S(\phi_2)$,
\be
\label{Deltaeta}
\Delta S = 4\chi^2\sqrt{\frac{N_{\gamma_1}(N_{\rm pass}+1)}{2N_{\gamma_2}}}
\left|
\sin\frac{lm^2}{4\omega}
\right| F(\phi_1,\phi_2) \,,
\ee
where
\bea
\label{f}
 F(\phi_1,\phi_2) &\equiv& 
 m^2
  \left|{\rm Re}\frac{e^{-i\phi_1}-e^{-i\phi_2}}{m^2 + \omega^2 \Delta\varepsilon_{\rm r}}\right|\nonumber\\
&=&\frac{m^2}{{{\left( {{m}^{2}}+\alpha \right)}^{2}}+{{\beta }^{2}}}
\big| {\left( {{m}^{2}}+\alpha \right)\left( \cos {{\phi }_{1}}-\cos {{\phi }_{2}} \right)}
\nonumber\\&&
-{\beta \left( \sin {{\phi }_{1}}-\sin {{\phi }_{2}} \right)}\big|\,. 
\eea
Here we have defined $\alpha=\omega^2{\rm Re}\Delta \varepsilon_{\rm r}$ and $\beta=\omega^2{\rm Im}\Delta \varepsilon_{\rm r}$. 

Note that the relative signal (\ref{Deltaeta}) depends on the ratio $N_{\gamma_1}/N_{\gamma_2}$ and, thus, it is maximized when $N_{\gamma_1} \gg N_{\gamma_2}$. Thus, the reference beam $\gamma_2$ must be much weaker than the main beam $\gamma_1$. However, the parameter $N_{\gamma_2}$ should not be vanishing as the interference term in Eq.\ (\ref{prob}) is absent in this case. In what follws we will assume that $N_{\gamma_2} \ll N_{\gamma_1}$ and $N_{\gamma_1}\approx N_\gamma$.

\subsection{Signal-to-noise ratio}

The relative signal (\ref{Deltaeta}) by itself does not allow one to make conclusions about the sensitivity of the proposed experimental setup. Indeed, the sensitivity of the experiment is specified by the requirement that the signal due to the dark photon detection should be greater than the (shot) noise in the photon beam $\gamma_2$. Therefore, it is important to estimate the signal-to-noise ratio (SNR).

Recall that the uncertainty in the number of photons in the beam is $\sqrt{N_{\gamma_2}}$, where $N_{\gamma_2}$ is the average number of photons in the beam described by the Poisson distribution. 
However, one can use the squeezed states of light to reduce this uncertainty. Although the practical implementation of such states is technically involved, this method allows to significantly reduce the noise in optical experiments based on interferometry, see e.g. \cite{Schnabel:2016gdi} for a review. In particular, the noise reduction by about 10 dB may be achieved with this technique. Therefore, we will assume that this noise reduction is applied in the dark photon detection experiment. Presumably, this will allow to reduce the shot noise in the detector down to $\frac13\sqrt{N_{\gamma_2}}$. Thus, the corresponding signal-to-noise ratio reads
\begin{equation}
    \begin{aligned}
  &{\rm SNR}\approx 3\eta \sqrt{{{N}_{{{\gamma }_{2}}}}}\\
  &=12{{\chi }^{2}}\sqrt{{{N}_{\gamma }}\frac{{{N}_{\text{pass}}}+1}{2}}\left| \sin \left( \frac{{{m}^{2}}l}{4\omega } \right) \right|F(\phi_1,\phi_2)\,. 
\end{aligned}
\end{equation}

The requirement that SNR should be greater than 1 allows one to impose the constraints on the coupling constant $\chi$,
\be
\chi^{-2}< 12\sqrt{{{N}_{\gamma }}\frac{{{N}_{\text{pass}}}+1}{2}}\left| \sin \left( \frac{{{m}^{2}}l}{4\omega } \right) \right|F(\phi_1,\phi_2)\,. 
\label{chi1}
\ee

The expression (\ref{chi1}) contains the function $F(\phi_1,\phi_2)$ defined in Eq.\ (\ref{f}). This function takes into account the refraction and absorption of photons in the detecting medium. Note that the atomic transitions in the detector are assumed to be in resonance with the laser light. Since the imaginary part of the refractive index is much larger than its real part near the atomic resonance, we can assume ${\rm Re}\Delta\varepsilon_{\rm r}\ll {\rm Im}\Delta\varepsilon_{\rm r}$. In this case, the function (\ref{f}) is near its maximum value at $-\phi_1=\phi_2=\pi/2$ (at least when the dark photon mass is small, $m^2< \omega^2 {\rm Im}\Delta\varepsilon_{\rm r}$)
\be
F_{\rm max} \approx \frac{2m^2 \omega^2 {\rm Im}\Delta \varepsilon_{\rm r}}{
(m^2+\omega^2{\rm Re}\Delta\varepsilon_{\rm r})^2
+\omega^4({\rm Im}\Delta \varepsilon_{\rm r})^2}\,.
\ee
Substituting this value into Eq.\ (\ref{chi1}) we obtain the sensitivity of the experiment
\be
\chi^{-2}< 
\frac{24\sqrt{{{N}_{\gamma }}\frac{{{N}_{\text{pass}}}+1}{2}}\left| \sin \left( \frac{{{m}^{2}}l}{4\omega } \right) \right|m^2 \omega^2 {\rm Im}\Delta \varepsilon_{\rm r}}{
(m^2+\omega^2{\rm Re}\Delta\varepsilon_{\rm r})^2
+\omega^4({\rm Im}\Delta \varepsilon_{\rm r})^2}\,.
\label{chi2}
\ee

When $m^2> \omega^2 {\rm Im}\Delta\varepsilon_{\rm r}$, the function (\ref{f}) reaches its maximum near the point $(\phi_1,\phi_2)=(0,\pi)$,
$F_{\rm max}\approx 2$.




\section{Numerical estimates}
\label{Num}

For numerical estimates, we assume that the laser produces $5.4\times 10^{21}$ photons per second for one day and that the resonant cavity allows for $N_{\rm pass}=20000$ reflections. 
The length $l$ of the photon beam may be of order $l=0.5$  m which is a reasonable length scale for a laboratory-based experiment.

For the target, in principle, it is possible to use any photon detector based on the atomic or molecular transitions in resonance with the given laser light. For simplicity of our estimates, as a detector we consider a dilute sodium vapor at temperature $T = 400\ {\rm K}$ and density $N_{\rm at}= 3.4\times 10^{10}\ {\rm cm}^{-3}$. The photon frequency $\omega=2.1022\ {\rm eV}$ is chosen to match that of the ${}^2S_{1/2} \rightarrow {}^2P_{1/2}$ transition. The real and imaginary parts of the relative permittivity are estimated in Appendix \ref{appx}, 
${\rm Re}\Delta\varepsilon_{\rm r} = 1.4\times 10^{-6}$, 
${\rm Im }\Delta\varepsilon_{\rm r} = 2.7\times 10^{-3}$. For this value of the relative permittivity the estimate (\ref{chi2}) applies. The corresponding exclusion region for the coupling constant $\chi$ is represented in Fig. \ref{ResultPlot} by the red line.

\begin{widetext}

\begin{figure}[h]
\begin{center}
\includegraphics[width=13cm]{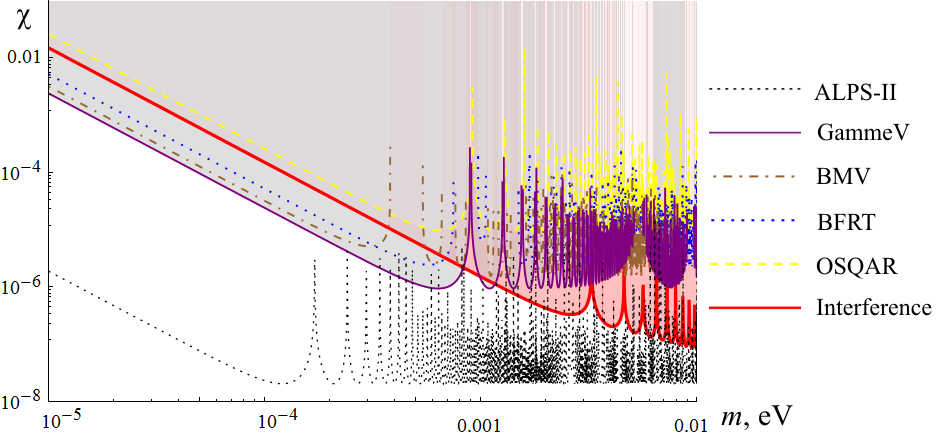}
\end{center}
\caption{Comparison of constraints on the dark photon coupling constant from different LSW-type experiments. Purple thin line corresponds to the GammeV experiment, brown dot-dashed line to the BMV experiment, blue dotted line to the BFRT experiment, yellow dashed line to the OSQAR experiment and red thick line represents possible constraints from the experimental setup proposed in this paper based on the interference between the photon and dark photon amplitudes. The short-dashed black line shows expected results from the ALPS-II experiment. The new (pink) exclusion region improves the constraints from other SLW experiments in the region of the dark photon masse $10^{-3}< m < 10^{-2}$ eV.}
\label{ResultPlot}
\end{figure}

\begin{table}[h]
\begin{tabular}{lccccccc}
\hline\hline
Experiment  &  $\omega$ [eV] & $N_{\rm pass}$ & $l_1$ [m] & $l_2$ [m] & $N_\gamma$ [Hz] & $\eta$ & $N_{\rm det}$ [Hz] \\\hline
ALPS-II & 1.17 & 5000 & 100 & 100 & $1.9\times 10^{20}$ & 0.75 & $10^{-6}$ \\ \hline
OSQAR & 2.41 & 1 & 7.15 & 7.15 & $3.9\times 10^{19}$ & 0.33 & 1.85\\ \hline
BFRT & 2.47 & 200 & 11 & 6.5 & $7.8\times 10^{18}$ & 0.055 & 0.018 \\ \hline
BMV & 1.17 & 1 & 20 & 1 & $6.7\times 10^{22}$  & 0.5 & 3.69 \\ \hline
GammeV & 2.3 & 1 & 5.4 & 7.2 & $6.6\times 10^{23}$ & 0.25 & 3.69\\
\hline\hline
\end{tabular}
\caption{Parameters of LSW-type experiments sensitive to the dark photon. Here $\omega$ is the frequency of the laser beam, $N_{\rm pass}$ is the number of passes through the cavity (if present), $l_1$ and $l_2$ are the lengths of the beam path before and after the wall, $N_\gamma$ is the photon  rate in the laser beam, $\eta$ is the quantum efficiency of the detector and $N_{\rm det}$ is the upper limit on the detected number of photons. The values of these parameters are extracted from the works 
\cite{ALPSII1,ALPSII2,Ballou:2015cka,Pugnat:2013dha,Pugnat:2007nu,Ahlers:2007rd}.
\label{tabl}}
\end{table}

\end{widetext}

In Fig.\ \ref{ResultPlot}, for comparison we collect also the results of other light-shining-through-a-wall  experiments: ALPS-II \cite{ALPSII1,ALPSII2}, GammeV \cite{Chou:2007zzc}, BMV \cite{Robilliard:2007bq}, BFRT \cite{Cameron:1993mr,Ruoso}, and OSQAR \cite{Ballou:2015cka,Pugnat:2013dha,Pugnat:2007nu}. These experiments are sensitive to dark photon particles if the coupling constant obeys 
\be
\chi > \frac12\left( \sin\frac{l_1 m^2}{4\omega} \sin\frac{l_2 m^2}{4\omega} \right)^{-\frac12}\left(\frac{N_{\rm det}}{\eta N_\gamma} \right)^{\frac14}
\,,
\ee
where $l_1$ and $l_2$ are the lengths of the beam paths before and after the wall, $\eta$ is the quantum efficiency of the detector and $N_{\rm det}$ is the upper limit on the detected number of photons. These parameters are collected in Table \ref{tabl}.

As is seen in Fig.\ \ref{ResultPlot}, the proposed experimental setup based on the interference of the photon and dark photon absorption amplitudes may provide new constraints on the dark photon coupling constant in the mass region $10^{-3}< m < 10^{-2}$ eV as compared with current constraints from the LSW experiments.

\section{Discussion}
\label{Disc}

In this paper, we investigated the possibility of using atomic transitions for detecting dark photon particles in the LSW-type experiments. The main advantage of this technique is that one can detect the dark photon signal which arises from the interference term for the photon and dark photon absorption amplitudes. As a result, the absolute signal becomes first order in the dark photon coupling constant. Analogous technique has recently been proposed for detecting axion particles \cite{Tan:2018tcs}. 

The proposed experimental setup may be considered as an alternative to the standard way of detecting dark photons (and axions) in the LSW experiments in which the non-transparent wall completely blocks the photon beam and the detection is limited by the sensitivity of the photon detector. If we assume that the photon beam is only partially blocked so that some (small) fraction of the photon beam reaches the detector, the detection is based on the excess of the signal due to interference between the photon and dark photon absorption amplitudes. This technique does not require very sensitive photon detectors such as those used in the ALPS-II experiment \cite{ALPSII1}. Instead, the detector should employ atomic transitions and resonantly absorb both photons and dark photons. 

The sensitivity of the proposed experimental setup is limited mainly by the shot noise in the photon beam which grows as the square root of the number of absorbed photons. However, the shot noise may be significantly reduced by applying the squeezed stated of light (see, e.g., \cite{Schnabel:2016gdi} for a review). In our case, we assumed that this technique may reduce the shot noise by 10 dB. This allows us to expect that the proposed experimental setup may improve existing constraints on the dark photon coupling constant in the mass region $10^{-3}< m < 10^{-2}$ eV. 

It is also important to note that the presence of any medium affects the oscillations between photons and dark photons. As was demonstrated in Sect. \ref{Sec-osc}, the stationary states of photons and dark photons in a medium are different from those in the vacuum. This significantly reduces the probability of dark photon oscillation. Therefore, to enhance the production of dark photons it is necessary to use a good vacuum on the course of laser beam. However, as was demonstrated in \cite{An:2013yfc,An:2013yua,An:2014twa}, in the dark photon detector the medium effects play also important role and reduce the signal. In Eqs.\ (\ref{Deltaeta}) and (\ref{chi1}) these effects are taken into account by the function $f$ which depends on the  relative  permittivity of the detecting medium as in Eq.\ (\ref{f}).

Finally we stress that the expected constraints on the dark photon coupling constant from the LSW experiments should be considered as complimentary to and independent from the results of other experiments devoted to the searches of dark matter particles and, in particular, dark photons. In particular, for mass $m$ below $10^{-4}$ eV the constraints from the experiments \cite{Williams:1971ms,Bartlett:1988yy} searching for the deviations from the Coulomb law appear to give very strong constraints on the dark photon coupling. As is shown in \cite{Redondo:2008aa}, the region $m>1$ eV is strongly constrained by the CAST experiment detecting axions and dark photons of solar origin. 

\subsection*{Acknowledgments}

We are grateful to Dmitry Budker for useful discussions. 
This work is supported by the Australian Research Council
Grant No. DP150101405 and by a Gutenberg Fellowship.

\appendix
\section{Refractive index of the detecting medium}
\label{appx}
As is argued in Sect.\ \ref{Sec-osc}, in considering the atomic transitions due to the dark photon absorption it is necessary to take into account the medium effects. We will assume that the detector is given by a gas with refractive index $n = \sqrt{\varepsilon_{\rm r}}$, where $\varepsilon_{\rm r}$ is the relative permittivity. In this section we will estimate this quantity for Sodium vapor at temperature $T=400$ K and pressure $P = 1.85\times 10^{-4}$ Pa. The corresponding atom density is $N_{\rm at} = 
3.35\times 10^{10}$ cm$^{-3}$.

We assume that the frequency of the laser light $\omega$ is in resonance with the $s_{1/2}\to p_{1/2}$ transition between the ground state and first excited state in Na,
$\omega=\omega_0 = 16956.17025$ cm$^{-1}$. The corresponding E1 reduced matrix element is
$A_{s_{1/2}\to p_{1/2}} \approx 3.5 a_{\rm B}$, where $a_{\rm B}$ is the Bohr radius. The nearest off-resonance E1 transition is $s_{1/2}\to p_{3/2}$ with the energy $\omega_1 = 16973.36619$ cm$^{-1}$. The corresponding E1 reduced matrix element reads $A_{s_{1/2}\to p_{3/2}} \approx 5.0 a_{\rm B}$.

The relative permittivity may be expressed via the atomic polarizability $\alpha(\omega)$,
\be
\Delta \varepsilon_{\rm r} \equiv n^2 -1 =4\pi N_{\rm at} \alpha(\omega)\,.
\ee
Near resonance, the atomic polarizability is given by 
\be
\alpha(\omega)= |A|^2 \left\langle 
\frac1{\omega - \omega_0 +i\Gamma/2}
\right\rangle\,,
\ee
where $A$ is the atomic transition matrix element, $\Gamma$ is the natural width and $\langle \ldots \rangle$ means the average over the thermal distribution of atoms in the target. Recall that the most probable speed of atoms of the gas with the temperature $T$ reads $v_0=\sqrt{2k_{\rm B}T/M}$, where $k_{\rm B}$ is the Boltzmann constant and $M$ is the atomic mass. This thermal motion of atoms creates the Doppler broadening of spectral lines. As is demonstrated in \cite{khrip}, after averaging, the real and imaginary part of the relative permittivity are
\bea
{\rm Re}\Delta\varepsilon_{\rm r}&=&4\pi N_{\rm at} |A_{s_{1/2}\to p_{1/2}}|^2
\frac{g(u,v)}{\Delta_D}\,,\\
{\rm Im}\Delta\varepsilon_{\rm r}&=&4\pi N_{\rm at} |A_{s_{1/2}\to p_{3/2}}|^2
\frac{f(u,v)}{\Delta_D}\,,
\eea
where $\Delta_D = \omega_0 v_0$, $u=\frac{\omega-\omega_0}{\Delta_D}$, 
$v = \frac{\Gamma}{2\Delta_D}$, and the functions $f$ and $g$ are defined as follows
\bea
f(u,v)&=& {\rm Re}\sqrt\pi e^{-(u+i v)^2}
 [1-{\rm erf}(v-iu)]\,,\\
g(u,v)&=& {\rm Im}\sqrt\pi e^{-(u+i v)^2}
 [1-{\rm erf}(v-iu)]\,.
\eea

For the sodium vapor under consideration we have $v_0 = 1.8 \times 10^{-6}$,
$\Delta_D = 3.8\times 10^{-6}$ eV and $v = 5.3\times 10^{-6}$. In this case, 
$f(0,v) = 1.8$ and 
\be
{\rm Im }\Delta\varepsilon_{\rm r} = 2.7\times 10^{-3}\,.
\ee
Note that the imaginary part of the relative permittivity defines the absorption length
$L=(\omega {\rm Im }\Delta\varepsilon_{\rm r})^{-1} = 0.2$ mm. Thus, the target should not be of large size.

To find the real part of the relative permittivity it is necessary to consider the contribution from the $s_{1/2}\to p_{3/2}$ transition. In this case, $v= (\omega_1 - \omega_0)/\Delta_D = 567$ and
$g(u,v) = 1.8\times 10^{-3}$. Thus, the real part of the relative permittivity is 
\be
{\rm Re}\Delta\varepsilon_{\rm r} = 1.4\times 10^{-6}\,.
\ee

\end{document}